\documentclass[twocolumn,10pt]{IEEEtran}
\usepackage{amsmath,amsfonts,amssymb,mathrsfs}
\usepackage{cite}
\usepackage{verbatim}
\usepackage{bm,bbm}
\usepackage{algorithm}
\usepackage{graphicx}
\usepackage{epstopdf}
\usepackage{subfigure}

\newtheorem{theorem}{Theorem}

\begin{document}
\title{Shifting Maximum Eigenvalue Detection in Low SNR Environment}

\author{Lin Zheng,~Robert C. Qiu~\IEEEmembership{Fellow,~IEEE}, ~Qing Feng, ~Xuebin Li% <-this % stops a space
\thanks{Lin Zheng is with the Department
of Information and Communication Engineering, Guilin University of Electronic Technology, Guilin, 541004,
China. (e-mail: gwzheng@gmail.com)}% <-this % stops a space
\thanks{Robert C. Qiu, Qing Feng, and Xuebin Li are with the Department of Electrical and Computer Engineering, Tennessee Technological University, Cookeville, TN 38505, USA. (e-mail:rqiu@tntech.edu)}% <-this % stops a space
%\thanks{This work was partly supported by NSF of China No. 61371107 and (US)
%NSF Grant No. CNS-1619250, the Foundation of Guangxi Broadband
%Wireless Communication \& Signal Processing Key Laboratory No. GXKL061501.}
%\thanks{Manuscript received ; revised January 11, 2018.}
}

%\markboth{Journal of \LaTeX\ Class Files,~Vol.~6, No.~1, January~2018}%
%{Shell \MakeLowercase{\textit{et al.}}: Bare Demo of IEEEtran.cls for Journals}

\maketitle

\begin{abstract}
%\boldmath
Maximum eigenvalue detection (MED) is an important application of random matrix theory in spectrum sensing and signal detection. However, in small signal-to-noise ratio environment, the maximum eigenvalue of the representative signal is at the edge of Marchenko-Pastur (M-P) law bulk and meets the Tracy-Widom distribution. Since the distribution of Tracy-Widom has no closed-form expression, it brings great difficulty in processing. In this paper, we propose a shifting maximum eigenvalue (SMED) algorithm, which shifts the maximum eigenvalue out of the M-P law bulk by combining an auxiliary signal associated with the signal to be detected. According to the random matrix theory, the shifted maximum eigenvalue is consistent with Gaussian distribution. The proposed SMED not only simplifies the detection algorithm, but also greatly improve the detection performance. In this paper, the performance of SMED, MED and trace (FMD) algorithm is analyzed and the theoretical performance comparisons are obtained. The algorithm and theoretical results are verified by the simulations in different signal environments.
\end{abstract}

\begin{IEEEkeywords}
maximum eigenvalue detection, random matrix theory, spectrum sensing.
\end{IEEEkeywords}

% \fi
\IEEEpeerreviewmaketitle

\section{Introduction}

\IEEEPARstart{W}{ith} the progress of high performance computing technology in recent years, the application of big data processing is more and more extensive. How to extract useful information efficiently has a higher requirement for dimensionality reduction of statistical data. The eigenvalue detection in the random matrix theory is a class of theory and method which efficiently reserves big data information and meet the low-complexity processing requirements. It is widely used in the fields of fault detection, spectrum sensing, radar signal detection and intrusion detection, and has attracted wide attention in recent years. In this paper, we study the eigenvalue detection in spectrum sensing, but the proposed method is a universal method based on random matrix theory, and it can also be applied to other related applications.

Spectrum sensing is the key technology of cognitive radio communication. By searching the spectrum hole and primary-slave user intelligent channel occupancy rules, cognitive radio achieves maximum spectrum efficiency. It resolves the conflict between the scarce frequency resources under high requirement of wideband communications and the low spectrum utilization in current spectrum allocation system. Early spectrum sensing methods include cyclostationarity detection, matched filtering, energy detection, and power spectrum density estimation. In recent years, with the development of random matrix theory, a number of detection methods based on characteristics of eigenvalues or empirical spectral have been studied intensively. Eigenvalue detection are mainly divided into two categories, one is based on single eigenvalue \cite{zeng2008,zeng2009}, and the other is based on multiple eigenvalues of signal covariance matrix \cite{lin2012,zhang2015,huang2015,cai2015,liu2017}.  In \cite{zeng2008}, Zeng first proposed the spectrum detection algorithm based on the maximum eigenvalue. Under the condition of a priori known noise variance, the Tracy-Widom distribution of the largest eigenvalue and the results of Tracy-widom distribution research are used to get better detection performance than traditional methods. \cite{zeng2009} eliminates the constraints requiring the prior known noise variance in the optimal threshold calculation in the MED algorithm through the ratio of the maximum and minimum eigenvalues (MME). More accuracy ratio distributions and thresholds for MME detection are derived in \cite{shakir2013} and \cite{zhou2018}, and in \cite{shakir2013} a generalized mean detectors based on the ratio of eigenvalues are introduced. However, in practical spectrum sensing, there often exist more than one primary user signal or multichannels in the detected frequency band, which causes multiple eigenvalues distribution in the empirical spectral. Therefore, it is a more reasonable way to combine multiple eigenvalues in detection. In \cite{lin2012}, the trace method of covariance matrix called FMD is applied to accumulate the eigenvalues, making full use of the multiple signal components in the empirical spectral to be measured, and achieves good detection performance in the DTV real signal environment. \cite{zhang2015} uses the central limit theorem in the random matrix theory \cite{qiu2017} to analyze the actual MIMO receiving signal through the eigenvalues accumulation which is in coincidence with the likelihood ratio detection form. Huang in \cite{huang2015} exploited a new spectrum detection based on the property that the ratio of the sum of eigenvalue higher moments to the higher moment of the eigenvalues sum approaches to a specific value, and the ratio is consistent with Gaussian distribution. \cite{cai2015} found and proved that logarithmic determinant of covariance matrix is consistent with the Gaussian distribution, which can also be used to signal detection. Liu in \cite{liu2017} obtained by theoretical derivation the optimal multi-eigenvalue combination weighting method in spectrum sensing. It solves the loss of detection performance caused by the cumulative variance of noise eigenvalue in the combination of all eigenvalues, such as trace method.

In addition, facing the inaccurate estimation with under sampling and sampling covariance matrix, \cite{lin2015} applies OAS covariance matrix estimation algorithm for cognitive radio in the primary user (PU) signal detection. In \cite{zeng2009cav} and \cite{jin2015}, the existence of the signal is directly determined by analyzing the difference between the covariance matrix of the primary user signal and the noise covariance matrix. Furthermore, the latter further improves the detection by adopting the optimal weighting method. Except for the methods from eigenvalue and covariance matrix property, traditional spectrum estimation has been improved for spectrum sensing. \cite{gao2014} and \cite{bomfin2017} estimate the power spectrum density through multi-band FFT multi-point energy cumulation in frequency domain. The presence of primary user signal in cognitve radio is determined by the ratio of the average subband power to the whole-band power. The latter letter further exploits the symmetrical double-side band property of real signal to improve detection performance. However, the methods based on multi-band power spectrum comparison are constrained by the non-uniform distribution of signal spectrum, thus the algorithm can not be used when the signal spectrum distribution is evenly distributed in the measured bandwidth.

The eigenvalue-based detection method is an important application of stochastic matrix theory in spectrum sensing and signal detection. Its advantage is better noise and signal separation than frequency domain based detection. However, the algorithm based on the maximum eigenvalue is constrained by Tracy-Widom distribution at the edge of M-P law bulk in low signal-to-noise ratio environment. The maximum eigenvalue distribution corresponding to the target signal is "compressed" at the bulk edge, so that the signal representation with the maximum eigenvalue is "buried". In this paper, we make full use of the different distribution characteristics of the maximum eigenvalue of the covariance matrix under different SNRs. Combining with the auxiliary signal with the same eigenvector as the signal to be detected, the hidden target eigenvalue at low signal-to-noise ratio is activated and breaks out from the boundary of the M-P law bulk, and its distribution changes from Tracy-widom distribution to Gaussian distribution. Therefore, the amplitude of the detection signal increases near linearly with the increase of the maximum eigenvalue. The detection efficiency is obviously improved without increasing the complexity of the algorithm.

The remainder of the paper is organized as follows: the second section gives the signal model and the necessary mathematical assumptions. In the third section, the proposed shifting maximum eigenvalue detection (SMED) is introduced. Prior to this, based on the recent progress on the maximum eigenvalue distribution in multivariate statistical theory in recent years, the distribution characteristics of maximum eigenvalues are analyzed in detail, which is the theoretical basis of the SMED algorithm. The fourth section analyzes and compares the performance of the two maximum eigenvalue detection: SMED and MED, as well as a kind of eigenvalues combination based detection, FMD. The differences between two classes of detection algorithms based on eigenvalue and eigenvalues accumulation are analyzed, and the influence of the detection environment to algorithms is presented. The fifth section carries on the algorithm simulations, verified the algorithm performances, and compared the performances of serveral kinds of detection algorithms under BPSK signal and DTV signal environment respectively.

\section{Signal Model}
One of the advantages of signal detection methods based on random matrix theory is its combinability based on eigenvalues or empirical spectral distribution (ESD). With this characteristic, whether multi-antenna signals or multi-sensor distributed sampling, combined signal detection does not require accurate synchronization. Therefore, the performance of signal detection actually still depends on the performance of single-channel signal detection algorithm. This paper is based on the more universal single-channel signal model, which can be further extended to multi-antenna or distributed signal processing environment. Suppose the oversampling frequency is adopted, and the received sampling signal is under $\mathcal{H}_1$  hypothesis and  $\mathcal{H}_0$ hypothesis according to the existence of the target signal component. The two hypothetical signal models is given by
\[
\mathcal{H}_0:x[k]=e[k]
\]
\[
\mathcal{H}_1:x[k]=h[k]s[k]+e[k]
\]
We make the following assumptions:
\begin{itemize}
\item (AS1) $e[k] $ is the complex white noise component of the received signal $x[k] $, and assumes that each sample $e[k] $  is independently and identically distributed (iid), with zero mean and variance $\sigma_e^2 $.
\item (AS2) $s[k] $ is the signal component in the received signal to be detected. It is the complex signal sampled by IQ channels.  Under $\mathcal{H}_1 $, the complex signal $s[k] $  has the amplitude distribution uncertainty after passing through the channel $h[k] $. Assume $|s[k]|^2=1 $, and $s[k] $  is independent with the noise $e[k] $. There are just one eigenvalue and eigenvector corresponding to $s[k] $.
\item (AS3) $h[k]$ is the channel coefficient with slowly varying characteristics, and it is independent with $s[k] $  and $e[k] $. In the sampling, $h[k]$ can be viewed as unchanged, that is $h[k]\approx h $.
\end{itemize}

According to the sampling time sequence, the received sampling signal is divided into $P$ packets, each packet has $N$ time-sequential sampling values. Thus, we have the $p$th sampling vector $\mathbf{x}_p= [x[(p-1)N+1],x[(p-1)N+2,\cdots,x[pN]]^T$. In another scenario, taking group according to $P$ receiving antennas, we have $N$ time-sequential samples in a group from the received antenna. The $p$th sampling vector is expressed by $\mathbf{x}_p=[x_p[1],x_p[2],\cdots,x_p[N]]^T $. Define $\mathbf{X}=(\mathbf{x}_1,\mathbf{x}_2,\cdots,\mathbf{x}_P)$. The statistics covariance matrix is given by
\begin{equation}
\mathbf{R}_x=\mathbb{E}[\mathbf{XX}^H]
\end{equation}
where $H$ is the conjugate operator.  When there is the signal to be detected in $x[k] $, i.e. under $\mathcal{H}_1 $ , there is
\begin{equation}
\begin{split}
\mathbf{R}_x &=h^2\mathbb{E}[\mathbf{SS}^H]+\mathbb{E}[\mathbf{EE}^H] \\
&=h^2\mathbb{E}[(\mathbf{s}_1,\cdots,\mathbf{s}_P)(\mathbf{s}_1,\cdots,\mathbf{s}_P)^H]+\sigma_e^2\mathbf{I}_N \\
&=h^2\mathbf{R}_s+\sigma_e^2\mathbf{I}_N
\end{split}
\end{equation}
Obviously, without signal to be detected in $x[k] $, i.e. under $\mathcal{H}_0 $, there is $\mathbf{R}_x=\sigma_e^2\mathbf{I}_N $. The eigenvalue distribution concentrated around $\sigma_e^2 $. Substitute the statistical covariance matrix with the sampling covariance matrix, which is defined as
\begin{equation}
\hat{\mathbf{R}}_x=\frac{1}{P}\mathbf{XX}^H=\frac{1}{P}\sum_{p=1}^P\mathbf{x}_p\mathbf{x}_p^H
\end{equation}

When $P$ is large enough, the eigenvalue distribution of the sampling covariance matrix, called as the empirical spectrum distribution (ESD), is asymptotically consistent with Marchenko-Pastur (M-P) law \cite{qiu2017}. Through the research in the field of multivariate statistical analysis in recent years, we have obtain a deep understanding of the distribution law of the maximum eigenvalue.

\section{Eigenvalues based Signal Detection}
The maximum eigenvalue detection is the classical algorithm based on random matrix theory. Its principle is based on Tracy-Widom distriubtion of the maximum eigenvalue of sample covariance matrix in small signal-to-noise ratio (SNR) environment. However, the Tracy-Widom distribution has no closed-form expression at present, and the distribution of eigenvalues near the edge of M-P law bulk shows a "compressed" property, which make it difficult to detect signals. In the following, we first analyze the distribution of maximum eigenvalues in $\mathcal{H}_1 $  hypothesis in detail and then give a way to get better detection performance by shifting the maximum eigenvalue by our designed auxiliary signal.

\subsection{Maximum Eigenvalues Theory}
The eigenvalues of sample covariance matrix $\hat{\mathbf{R}}_x $  are represented by $\lambda\in\{\lambda_1,\lambda_2,\cdots,\lambda_N\}$. Let $\lambda_1\ge\lambda_2\ge\cdots\ge\lambda_N $, and $\lambda_1 $ is the maximum eigenvalue. From the Marchenko-Pastur law \cite{qiu2017}, when the elements in $\mathbf{X} $  are independently and identically distributed (iid), zeros mean and variance $\sigma_e^2 $ ，with $N\rightarrow \infty $  and $P\rightarrow \infty $, the eigenvalue $\lambda $ converges in the distribution given by
\begin{equation}
f^{MP}(\lambda)=\begin{cases}
\frac{1}{2\pi c\sigma_e^2 \lambda}\sqrt{(\lambda_1-\lambda)(\lambda-\lambda_N)} & a\leq \lambda\leq b \\
0, & \text{else}
\end{cases}
\end{equation}
where $c=N/P (c<1)$, $\lambda_N=\sigma_e^2(1-\sqrt{c})^2 $, $\lambda_1=\sigma_e^2(1+\sqrt{c})^2 $. In 2005, J. Baik in \cite{baik2005} firstly gave the systematic analysis about the maximum eigenvalue distribution. It has below key results.

Let $\ell_1,\cdots,\ell_N $  respresent the eigenvalues of statistical covariance matrix with normalized noise components $(\sigma_e^2=1) $, and $\ell_1\ge\ell_2\ge\cdots\ge\ell_N $.

\begin{theorem}  \label{thm:thm1}
Without signal to be detected in $x[k] $, i.e. null case, there is $\mathbf{R}_x=\mathbf{I}_N $ , that is $\ell_{1,\cdots,\ell_N}=1 $; Otherwise, when $1<\ell_{1,\cdots,r}\leq 1+\sqrt{c} $ , $\ell_{r+1,\cdots,N}=1 $($k<r<N $ ), the distribution of $\lambda_1 $, the maximum eigenvalue of $\hat{\mathbf{R}}_x $ ,  follows Tracy-Widom distribution.
\begin{equation}
\mathbb{P}\left((\lambda_1-(1+\sqrt{c})^2)\cdot\frac{c^{-1/2}}{(1+c^{-1/2})^{4/3}}P^{2/3}\leq x\right)\rightarrow F_k(x)
\end{equation}
where $F_k(x) $  is the Tracy-Widom distribution function, and the value of $k$ is given by
\begin{equation}
F_k(x)=\begin{cases}
F_0(x),& 0<\ell_{1}<1+\sqrt{c}\\
F_1(x),& \ell_k<1+\sqrt{c}=\ell_1 \\
F_2(x),& \ell_k<1+\sqrt{c}=\ell_{1,2}\\
\cdots & 1+\sqrt{c}=\ell_{1,2,\cdots,r}
\end{cases}
\end{equation}
\end{theorem}

At the same time, when $\mathbf{X} $  is the complex Gaussian random signal, and its covariance is random complex Hermitian matrix from the Gaussian unitary ensemble (GUE), there is $F_{GUE}=F_0(x) $  at null case. When $\mathbf{X} $  is the real Gaussian random signal, and its covariance is random real Hermitian matrix from the Gaussian orthogonal ensemble (GOE), there is $F_{GOE}=F_1(x) $ at null case. In recent \cite{chiani2014}, a simple and accurate method to approximate Tracy-Widom distribution with shifted gamma distribution function was given.

\begin{theorem} \label{thm:thm2}
when there is signal to be detected in $x[k] $, and $\ell_{1,\cdots,k}> 1+\sqrt{c},(k\leq r) $ , $\lambda_1 $  follows Gaussian distribution
\begin{equation}\label{eq:thm2}
\mathbb{P}\left(\Big(\lambda_1-(\ell_1+\frac{\ell_1 c}{\ell_1-1})\Big)\cdot\frac{\sqrt{\beta P}}{\ell_1\sqrt{1-\frac{c}{(\ell_1-1)^2}}}\leq x\right)\rightarrow G_k(x)
\end{equation}
where $G_k(x) $  is the Gaussian distribution function. At $\ell_1>1+\sqrt{c}>\ell_2 $ , $G_k(x)=G_1(x)=\frac{1}{\sqrt{2\pi}}\int_{-\infty}^x e^{-\xi^2/2}d\xi $.
\end{theorem}

\begin{figure*}[t]\center
\includegraphics[width=0.8\textwidth, height=0.2\textheight]{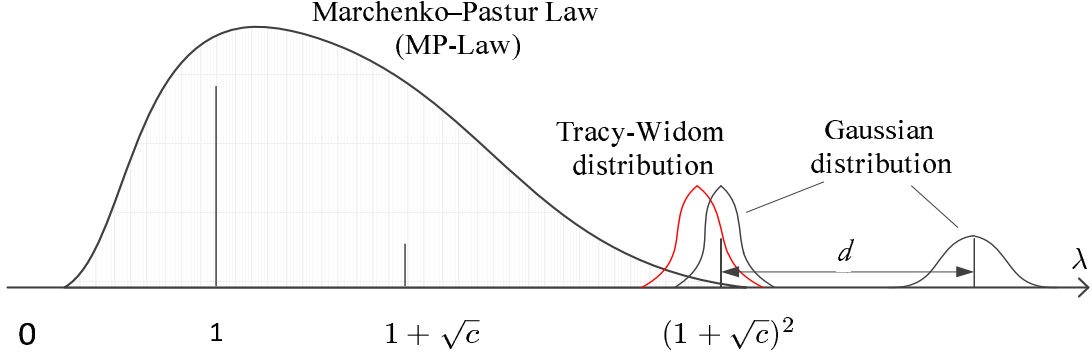}
\caption{Distribution of maximum eigenvalue of sample covariance matrix}\label{fig:mplaw}
\end{figure*}

Figure \ref{fig:mplaw} shows the distribution of maximum eigenvalues. It is noteworthy that $\beta=1 $  with GUE, and to GOE, there are different variances when $\ell_1 $  near $1+\sqrt{c} $. In \cite{feral2009,wang2009,mo2012}, $\beta $  is respectively 1, 2, 1/2.

When the signal to be detected exists in the sample $x[k] $, and it causes the only and maximum eigenvalue, maximum eigenvalue $\ell_1 $  of the normalized statistical covariance matrix has the relationship with SNR
\[
\ell_1=\frac{NP_s+\sigma_e^2}{\sigma_e^2}=1+N\cdot SNR
\]
where $P_s=h^2|s[k]|^2=h^2 $ , the only eigenvalue from the signal component in $x[k] $  is $NP_s+\sigma_e^2 $ , and $P_s/\sigma_e^2 $  is the SNR of $x[k] $.  It can be seen from the above distribution properties of eigenvalues that $\lambda_1 $  randomly clusters around $(1+\sqrt{c})^2 $  and follows Tracy-Widom distribution with $1<\ell_1<1+\sqrt{c} $, i.e. low SNR condition. With $\ell_1 $  improving under $1<\ell_1<1+\sqrt{c} $ , the distribution of $\lambda_1 $  has no significant change and is just like being compressed around $(1+\sqrt{c})^2 $, which results in the difficult extraction of $\lambda_1 $  from M-P Law bulk. With $\ell_1>1+\sqrt{c} $, i. e. higher SNR condition, shown in Eq. \ref{eq:thm2}, $\lambda_1 $  follows Gaussian distribution and increases with $\ell_1 $  at approximately linear. It is obviously more suitable for signal detection.
% figure --------------------------------------------------------
%figure ----------------------------------------------------------

\section{Shifting Maximum Eigenvalue Detection}
Based on the above analysis, under the conditions of low signal-to-noise ratio, i.e. $\ell_1<1+\sqrt{c} $ , the maximum eigenvalue in MED detection meets Tracy-widom distribution limit at the M-P law bulk edge. In this paper, we propose a method based on auxiliary signal, which has the same eigenvector with the component to be detected in received signal. Obviously, the eigenvalue   corresponding to the eigenvector is increased for the combined signal. When $\ell_1>1+\sqrt{c} $ , $\lambda_1 $ will be shifted right out of the bulk edge shown as in Fig. \ref{fig:mplaw}. At this time, $\lambda_1 $ follows Gaussian distribution. Therefore, the proposed method is called shifting maximum eigenvalue detection (SMED).

In some detection applications, the target signal to be detected has the single eigenvalue and eigenvector. Without loss of generality, we assume that $s[k] $  just has single eigenvalue to simplify the analysis. Thus, in Theorem \ref{thm:thm1} and Theorm \ref{thm:thm2}, there is $r=1 $ and $\ell_1>1=\ell_2=\cdots,=\ell_N $ .

The added auxiliary signal is represented by $s^v[k] $ , and its matrix is $\mathbf{S}^{v}=(\mathbf{s}^v_1,\cdots,\mathbf{s}^v_P)\in\mathbb{C}^{N\times P} $ . The combined signal of the received signal and the auxiliary signal has the statistical covariance matrix $\mathbf{R}^v_s=\mathbb{E}[h\mathbf{S}+h_v\mathbf{S}^v)(h\mathbf{S}+h_v\mathbf{S}^v)^H] $ , where $h_v $  is the amplitude of the auxiliary signal. Without a $priori$ phase information of the signal to be detected, the auxiliary signal is not phase synchronized. Therefore, the variance of the combined signal is $E_s^v=h^2+h_v^2 $ , and $\mathbf{R}_x^v=\mathbf{R}_s^v+\sigma_e^2\mathbf{I}_N $. The maximum eigenvalue $\ell_1 $  has
\begin{equation}
\ell_1=1+\frac{N(P_s+P_v)}{\sigma_e^2}=1+\frac{N(h^2+h_v^2)}{\sigma_e^2}=1+\gamma
\end{equation}
where $\gamma $  can be viewed as signal-to-noise ratio (SNR).  Since $\ell_1>1+\sqrt{c} $  after combining the auxiliary signal, the maximum eigenvalue of sample covariance matrix $\lambda_{\max} $ , i.e. $\lambda_1 $ , is in Gaussian distribution according to Theorem \ref{thm:thm2}. Its mean and variance are given by
\begin{equation}\label{eq:gaumean}
\begin{split}
\bar{\lambda}_{\max}&=\ell_1+\frac{\ell_1 c}{\ell_1-1}=(1+\gamma)+\frac{(1+\gamma)c}{1+\gamma-1}\\
&=(1+\gamma)\left(1+\frac{c}{\gamma}\right)
\end{split}
\end{equation}
\begin{equation}\label{eq:gauvar}
\begin{split}
\nu^2=\text{Var}(\lambda_1)&=\frac{1}{P}\left(\ell_1^2-\frac{c\ell_1^2}{(\ell_1-1)^2}
\right)\\
&=\frac{1}{P}(1+\gamma)^2\left(1-\frac{c}{\gamma^2}\right)
\end{split}
\end{equation}

In the shifting maximum eigenvalue detection (SMED), the constant false alarm probability is expressed as
\begin{equation}\label{eq:contpfa}
\begin{split}
P_{fa}&=P\left(\frac{\lambda_{\max}}{\bar{E}_e}>\zeta\Big|\mathcal{H}_0\right)\\
&=P\left(\lambda_{\max}>\zeta \bar{E}_e\right) \\
&=P\left(\frac{\lambda_{\max}-\bar{\lambda}_{\max}^0}{\nu_0}>\frac{\zeta N\sigma_e^2-\bar{\lambda}^0_{max}}{\nu_0}\right)\\
&=Q\left(\frac{\zeta N\sigma_e^2-\bar{\lambda}^0_{max}}{\nu_0}\right)\\
&\Longrightarrow \zeta
\end{split}
\end{equation}
where $\bar{E}_e $  means noise energy, $\zeta $  is the optimal threshold of constant false alarm detection, $\bar{E}_e=N\sigma_e^2 $  can be approximated by $\text{Tr}(\hat{\mathbf{R}}_x) $ . Under $\mathcal{H}_0 $  hypothesis,   $\bar{\lambda}_{max}^0 $ denotes the mean of maximum eigenvalue at null case. $\nu_0$  is the standard variance of ${\lambda}_{max}^0 $ . Combined with the auxiliary $\mathbf{S}^v $, the maximum eigenvalue of $\mathbf{R}_x^v $  satisfies $\ell_1^0=1+Nh_v^2/\sigma_e^2=1+\gamma_0>1+\sqrt{c} $. Thus, we have $\lambda_{\max}^0=\lambda_1^0\sim\mathcal{N}(\bar{\lambda}_{\max}^0,\nu_0^2) $. From Eq. \ref{eq:gaumean} and Eq. \ref{eq:gauvar}, we have $\bar{\lambda}_{max}^0=(1+\gamma_0)(1+c/\gamma_0) $, and $\nu_0^2=(1+\gamma_0)^2(1-c/\gamma_0^2)/P $. In Eq. \ref{eq:contpfa}, the Q function is defined as
\[
Q(x)=\frac{1}{\sqrt{2\pi}}\int_x^{+\infty}e^{-u^2/2}du
\]
When the constant $P_{fa} $  is assigned, we have
\begin{equation}\label{eq:pfadet}
\zeta=\frac{Q^{-1}(P_{fa})\nu_0+\bar{\lambda}^0_{\max}}{N\sigma_e^2}
\end{equation}

Next, the amplitude of the auxiliary signal $h_v $  is to be determined.  The value of $h_v $  should improve the detection performance. In radar signal detection, there is the detection SNR given by $SNR=(\mathbf{S}_1-\mathbf{S}_0)\mathbf{\Sigma}^{-1}(\mathbf{S}_1-\mathbf{S}_0)^T $ , where $\mathbf{S}_1 $  is $\bar{\lambda}_{\max}=\bar{\lambda}_1 $  , $\mathbf{S}_0 $  is $\bar{\lambda}^0_{\max}=\bar{\lambda}_1^0 $ , $\lambda_1^0 $  and $\lambda_1 $  is scalar. Thus, the covariance matrix $\mathbf{\Sigma} $  corresponds to $\nu^2 $ . Thus, there is
\begin{equation}\label{eq:snrd}
SNR=\frac{(\bar{\lambda}_{\max}-\bar{\lambda}^0_{\max})^2}{\nu^2}
\end{equation}

To improve the detection SNR,  the variance $\nu_0 $  of $\lambda_1^0 $  and the variance $\nu $  of $\lambda_1 $  should be kept as small as possible. The distance $d=\bar{\lambda}_{\max}-\bar{\lambda}^0_{\max} $  defined in Eq. \ref{eq:snrd} should be kept as large as possible. By analyzing the relationship between $SNR=d^2/\nu^2 $  and $\ell_1 $ , $h_v $  is optimized. Defining the original SNR as $\gamma_s=NP_s/\sigma_e^2 $ , and from Eq. \ref{eq:gaumean}, we have
\begin{equation}
d(\ell_1)=\bar{\lambda}_{\max}-\bar{\lambda}^0_{\max}=\gamma_s-\frac{c}{\ell_1-\gamma_s-1}+\frac{c}{\ell_1-1} \\
\end{equation}

\begin{figure}[t]\center
\includegraphics[width=0.5\textwidth, height=0.3\textheight]{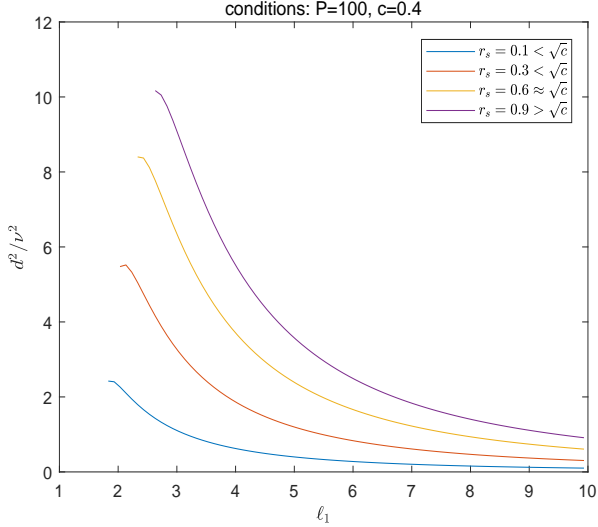}
\caption{curve of detection SNR $d^2/\nu^2\sim\ell_1=1+\sqrt{c}+\gamma_s+\Delta\ell $}\label{fig:vsopm}
\end{figure}

From $d(\ell_1)$ and Eq. \ref{eq:gauvar}, the curve of $d^2/\nu^2 $  according to $\ell_1 $  is illustrated in Fig.\ref{fig:vsopm}. Let $\ell_1=1+\sqrt{c}+\gamma_s+\Delta\ell$. We can see from the figure that $d^2/\nu^2 $  is degraded with the improved $\Delta\ell $ . It means that the amplitude of auxiliary signal should drive $\ell_1^0\rightarrow (1+\sqrt{c})^+ $  under   $\mathcal{H}_0 $ condition, where $(1+\sqrt{c})^+ $  denotes the minimum value larger than $1+\sqrt{c} $ .

Because that the $h_s $  of the signal to be detected is unknown, i.e. unknown $P_s $ , to ensure that the maximum eigenvalue   $\ell_1^0\rightarrow ( 1+\sqrt{c})^+ $ after combining with the correlated auxiliary signal $\mathbf{S}^v $, it can be deduced from the above conclusion that
\begin{equation}
P_v=\frac{\sigma_e^2(\sqrt{c}+\Delta\ell)}{N}
\end{equation}
% figure --------------------------------------------------------
%figure ----------------------------------------------------------

\section{Performance analysis and Comparison}
\subsection{SMED performance analysis}
Because the SNR of the detection signal has a definite relationship with the signal detection performance ROC, this paper analyzes the SNR and compares the performance of different detection algorithms. From (7), there is $\gamma=\gamma_v+\gamma_s$. According to Eq. \ref{eq:pfadet}, the detection SNR of SMED is given by
\begin{equation}\label{eq:smedperf}
\begin{split}
\text{SNR(SMED)}&=\frac{(\bar{\lambda}_{\max}-\bar{\lambda}^0_{\max})^2}{\nu^2}\\
&=\frac{P\left(\gamma_s-c(1/\gamma_v-1/\gamma)\right)^2}{(1+\gamma)^2(1-c/\gamma^2)}
\end{split}
\end{equation}
where $\gamma_v=NP_v/\sigma_e^2=(\sqrt{c}+\Delta\ell) $.

\subsection{MED performance analysis}
This paper analyzes the case of complex signal MED performance. Under the small SNR condition, the classical maximum eigenvalue detection (MED) exploits $\lambda_{\max} $ the characteristics of Tracy-Widom distribution.  Under $\mathcal{H}_0 $  hypothesis, the maximum eigenvalue follows Tracy-Widom distribution $\mathcal{TW}2 $  with the distribution function $F_{GUE}(x) $ , which is defined as $F_0(x) $  in \cite{baik2005} and defined as $F_2(x) $ in \cite{chiani2014}. There is no closed-form expression for Tracy-Widom distribution, and its distribution function $F_{GUE}(x) $ is defined as
\[
F_{GUE}=\exp\left\{-\int_x^{\infty}(y-x)q^2(y)dy\right\}
\]
where $q(y) $ is the solution of the nonlinear Painleve´ II differential equation: $q''(y)=yq(y)+2q^3(y) $. Its standard variance is given by
\[
std(\lambda_{\max}^0)=\frac{(1+\sqrt{c})^{4/3}}{P^{2/3}\sqrt{c}}
\]
The peak position of the probability density of $\mathcal{TW}2 $  distribution is a correction of its mean position:
\[
\bar{\lambda}_{\max}^0 =(1+\sqrt{c})^2-E_{\mathcal{TW}2}\cdot std(\lambda_{\max})
\]
where $E_{\mathcal{TW}2}= 1.771086807$.
With the signal to be detected and $\gamma_s>\sqrt{c} $ , $\lambda_{\max} $  follows Gaussian distribution. Its mean and variance has been given in Eq. \ref{eq:gaumean} and Eq. \ref{eq:gauvar}. At $0<\gamma_s<\sqrt{c} $ , the recently given Tracy-Widom distribution expression is not accurate as a function of the statistical eigenvalue $\ell_1 $ . Therefore, this paper just considers the case $\ell_1=1+\gamma_s>1+\sqrt{c} $. The SNR curve trend gives an approximate analysis of the unsolved performance. Taking the  $\lambda_1 $ variance under signal existed as the denominator of SNR, we have the following SNR results:
\begin{equation}\label{eq:medperf}
\begin{split}
&\text{SNR(MED)}=\frac{(\bar{\lambda}_{\max}^{(s)}-\bar{\lambda}^0_{\max})^2}{\nu^2} \\
& =\frac{P(1+\gamma)(1+c/\gamma)-(1+\sqrt{c})^2+1.771 P^{1/3}(1+\sqrt{c})^{4/3}/\sqrt{c}}{(1+\gamma)^2(1-c/\gamma^2)}
\end{split}
\end{equation}
where $\gamma=\gamma_s $ .

\subsection{FMD performance analysis}
Proposed in \cite{lin2012}, FMD signal detection applies the trace operation on the sampling covariance matrix. Let $T^0_x $  be the trace under no signal to be detected, $T_x $  be the trace under having signal to be detected. The decision method in FMD is   $T_x/T_x^0\gtrless^{\mathcal{H}_1}_{\mathcal{H}_0}\zeta $ where $\zeta $  is the optimal threshold. Due to being the combination of independently and identically distributed $\lambda_i $ , the distribution of the trace follows Gaussian distribution. Since $T_x=\text{Tr}(\hat{\mathbf{R}}_x)/N=\frac{1}{N}\sum_{n=1}^N\lambda_n $ , FMD is also an signal detection based on eigenvalue. Under $\mathcal{H}_0 $  hypothesis, there is the mean $\bar{T}_x^0=\sigma_e^2$ . Under $\mathcal{H}_1 $  hypothesis, the mean and variance of $T_x $  are given by
\begin{equation}
\begin{split}
\bar{T}_x &=E[\text{Tr}(R_s)/N]\\
&=\frac{1}{NP}E\left[\sum_{i=1}^N\sum_{n=1}^{P}(x_{i,n}+s_{i,n})^2\right]
=\sigma_e^2+P_s^2
\end{split}
\end{equation}

%figure ----------------------------------------------------------
\begin{figure*}[t]
\begin{equation}\label{eq:vartrace}
\begin{split}
&Var[T_x]=Var[\text{Tr}(R_s)/N]=\frac{1}{N^2P^2}E\left[\left(\sum_{i=1}^N\sum_{n=1}^{P}(x_{i,n}^2+s_{i,n}^2+2x_{i,n}s_{i,n})\right)^2\right]-(\sigma_e^2+P_s^2)^2 \\
&=\frac{1}{N^2P^2}\Big( NP E[x_{i,n}^4] + C_{NP}^2 2E[x_{i,n}^2 x_{j,m}^2]+NPE[s_{i,n}^4]+NP4E[x_{i,n}^2s_{i,n}^2]+ \\
&\qquad (NP)^2 2E[s_{i,n}^2x_{i,n}^2]+C_{NP}^2 2E[s_{i,n}^2s_{j,m}^2]\Big)-(\sigma_e^2+P_s^2)^2 \quad (i,n\neq j,m) \\
&=\frac{1}{N^2P^2}\Big( NP \cdot 2\sigma_e^4 + C_{NP}^2 2\sigma_e^4 + (NP)^22P_s^2\sigma_e^2+C_{NP}^22p_s^4+
NPE[s_{i,n}^4]+4NP\sigma_e^2P_s^2\Big)-(\sigma_e^2+P_s^2)^2 \\
&=\frac{\sigma_e^4+4\sigma_e^2P_s^2+E[s_{i,n}^4]-P_s^4}{NP}
\end{split}
\end{equation}
\hrulefill
\end{figure*}
%figure ----------------------------------------------------------

Approximate $E[s_{i,n}^4]/P_s^4\approx 0 $  and $1/N\approx 0 $ , we have
\begin{equation}\label{eq:fmdperf}
\begin{split}
&\text{SNR(FMD)}=\frac{(\bar{T}_x-\bar{T}_x^0)^2}{Var[T_x]}\\
&=\frac{P_s^4}{D(\text{Tr}(R_s)/L)}\approx \frac{P}{N/\gamma^2+4/\gamma}
\end{split}
\end{equation}

\begin{figure}[t]\center
\includegraphics[width=0.5\textwidth, height=0.3\textheight]{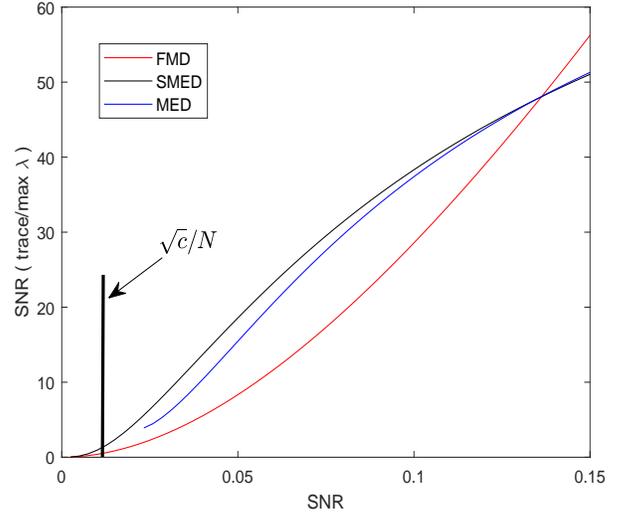}
\caption{Detection SNR curves of three detections to original SNR}\label{fig:snr2snr}
\end{figure}

According to Eq. \ref{eq:smedperf},\ref{eq:medperf}, and \ref{eq:fmdperf}, the detection SNR theoretic curves to the SNRs of the received signal, called original SNRs, are depicted in Fig. \ref{fig:snr2snr}. Shown in the figure, the detection performances of SMED and MED are better than that of the algorithm based on trace of sample covariance matrix. Only in the case of $SNR>0.14$, trace based algorithm shows better performance. The position of $\gamma=\sqrt{c} $  is signed in the figure. When $\ell_1 $ approaches to $1+\sqrt{c} $ , i.e. $SNR\approx\sqrt{c}/N $, current literatures have not given the accurate distribution of maximum eigenvalue at signal case. Therefore, the MED curve is calculated at the original SNR just slightly larger than $\sqrt{c}/N $. By shifting maximum eigenvalue, SMED avoids the uncertain distribution of $\lambda_{\max} $  near $SNR\approx\sqrt{c}/N $ . Overall, with the single eigenvalue and eigenvector of the signal to be detected, the order of the signal detection performances at low SNR is $SMED>MED>FMD$.

It is worth noting that SMED requires prior properties of the signal to be detected, and the signal has just one eigenvector. Therefore, SMED is more suitable for accurate searching of hidden signal with known single eigenvector. When the received signal has more than one eigenvalue and eigenvector, such as multiple primary user signals or broadband signal, SMED and MED and other eigenvalue detection class algorithms may be unsuitable for spectrum sensing. On the contrary, the algorithm based on combining eigenvalues, such as function of matrix based detection (FMD) \cite{lin2012}, optimal eigenvalues weighting detection (OEW)\cite{liu2017}, may be more suitable for spectrum sensing in a environment with mulitple primary user signals existing or wideband spectrum.

Suppose that there are $M$ eigenvalues in empirical spectral, and the total energy of the signal to be detected is the $\mathcal{M}-1 $  times of the $\lambda_{1} $  component, that is $\gamma=\sum_{i=1}^{M+1}\gamma_i=\mathcal{M}\gamma_1 $ . Obviously, from Eq. \ref{eq:fmdperf}, we have
\begin{equation}
\text{SNR(FMD)}= \frac{P\mathcal{M}^2}{N/\gamma_1^2+4\mathcal{M}/\gamma_1}
\end{equation}
% figure --------------------------------------------------------
%figure ----------------------------------------------------------
The trace-based detection performance improves about $\mathcal{M}^2 $  times than that in single eigenvalue case. The maximum eigenvalue detection algorithm can not get this gain. Therefore, detection application environment will also affect the performance of the algorithm.

\section{Simulations}
This section verifies the analyzed performances of the eigenvalue signal detection algorithms described above by simulations, illustrates the correctness of the theoretical results, and gives the performance comparison. More performance comparisons with recent detection algorithms are done in single eigenvalue environment and multiple eigenvalues environment. In addition, this section also gives the simulation results and analysis of SMED algorithm in terms of the amplitude and the correlationship of optimal auxiliary signal.

 % figure --------------------------------------------------------
 \begin{figure*}[t]\center
 \subfigure[SMED]{
 \includegraphics[width=0.322\textwidth, height=0.25\textheight]{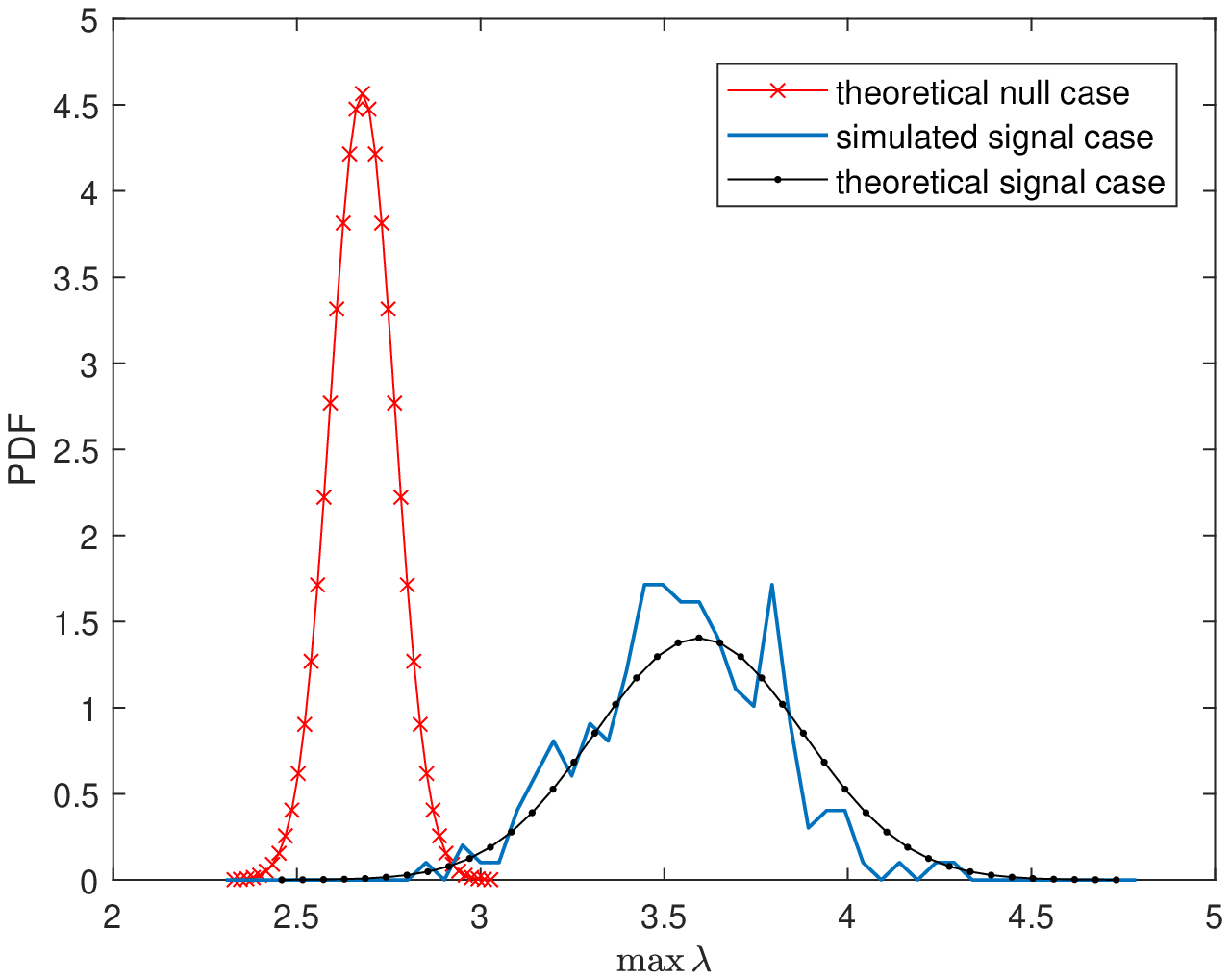}}
 \subfigure[MED]{
 \includegraphics[width=0.322\textwidth, height=0.25\textheight]{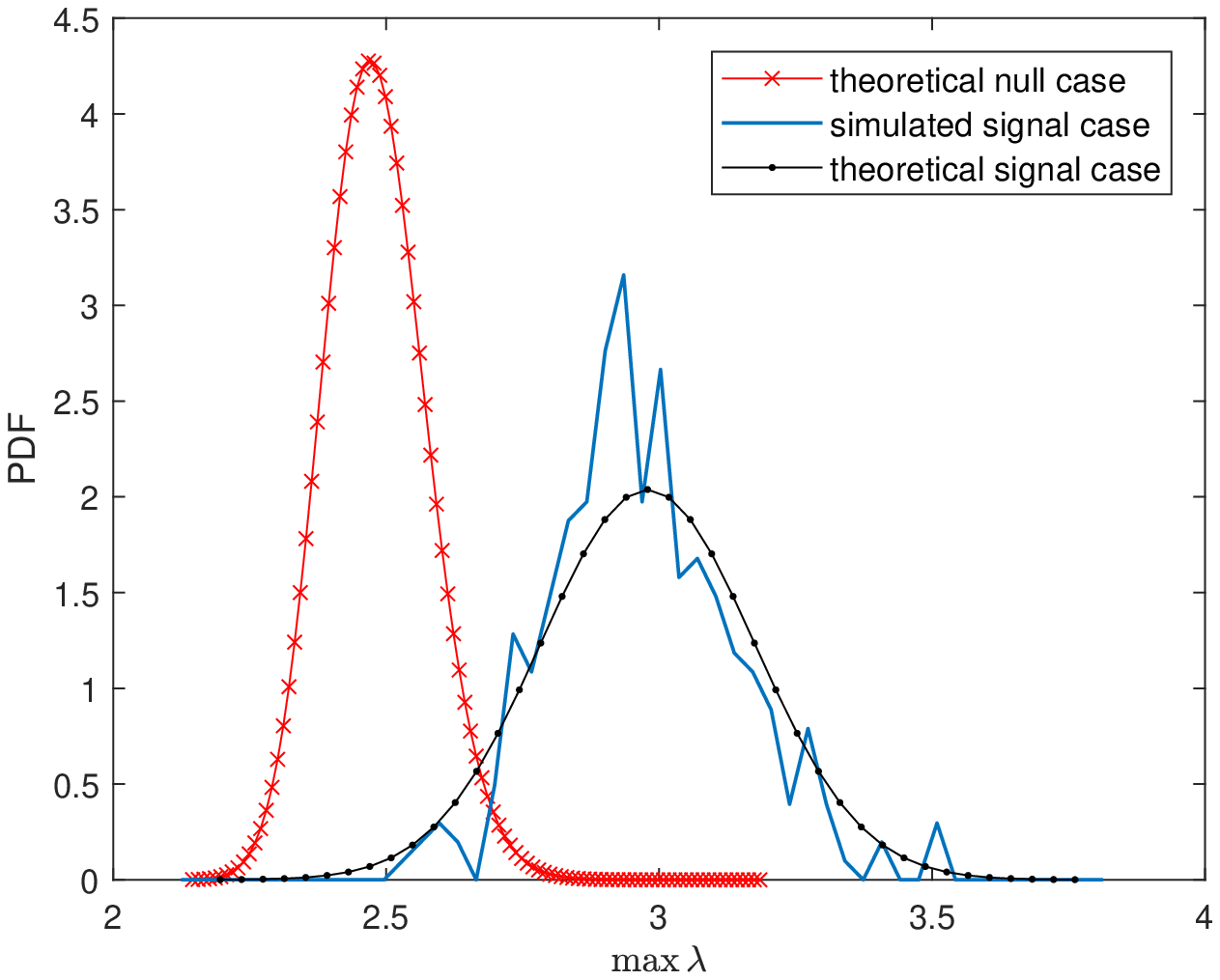}}
 \subfigure[FMD]{
 \includegraphics[width=0.322\textwidth, height=0.25\textheight]{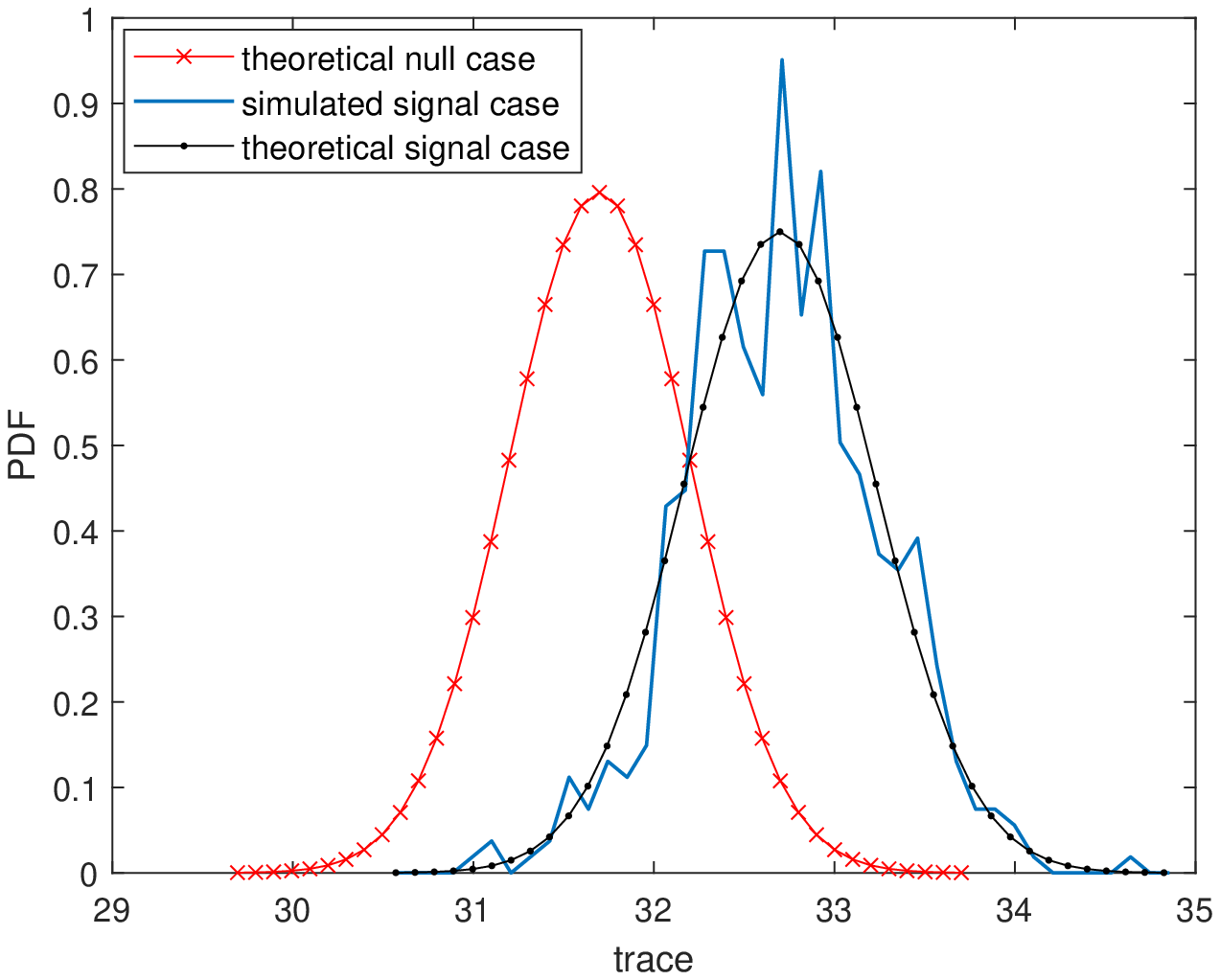}}
\caption{Probability density distribution of eigenvalue detections}\label{fig:dist}
 \end{figure*}
 % figure --------------------------------------------------------

Simulation uses BPSK test signal and complex Gaussian noise, the information rate of 100kbit/s, BPSK modulation frequency is 1MHz, oversampling frequency applies 10Msps. Take $N = 40$, $P = 100$, with $c = 0.4$, and the only maximum eigenvalue corresponding to the BPSK signal is undoubt. In the simulation experiment, the auxiliary correlated signals in SMED applies a continuous carrier frequency signal of 1MHz with random phase. At original SNR = -15dB of the received signal, it is shown in Fig. \ref{fig:dist} by simulation and theoretical calculations. The red curve on the left in the figure is the theoretical probability density function (PDF)  of maximum eigenvalue where there is only noise and no signal to be detected. The black and blue curves in the figure are respectively the theoretical and simulated PDF of the maximum eigenvalue in the case of detecting signal exists. The figure shows that the theory and simulations are consistent, verifying the correctness of the theoretical analysis.

 It is worth noting that the theoretical red curve of null case in Fig. \ref{fig:dist}(b) denotes $\mathcal{TW}2 $  distribution, while the maximum eigenvalue follows Gaussian distribution at signal case with $\ell_1=1+\gamma=1+N\cdot SNR=1+40\cdot 10^{-15/10}=2.265>1.633=1+\sqrt{c} $ . Under same SNR, the detection distance of SMED between the null case and the signal case is the largest one among three eigenvalue detections. Obviously, SMED has the best detection performance at the single signal eigenvalue and eigenvector environment. The performance followers are MED and FMD in order. It is also consistent with the theoretical results in Fig. \ref{fig:snr2snr}.

 % figure --------------------------------------------------------
\begin{figure}[t]\center
\includegraphics[width=0.5\textwidth, height=0.3\textheight]{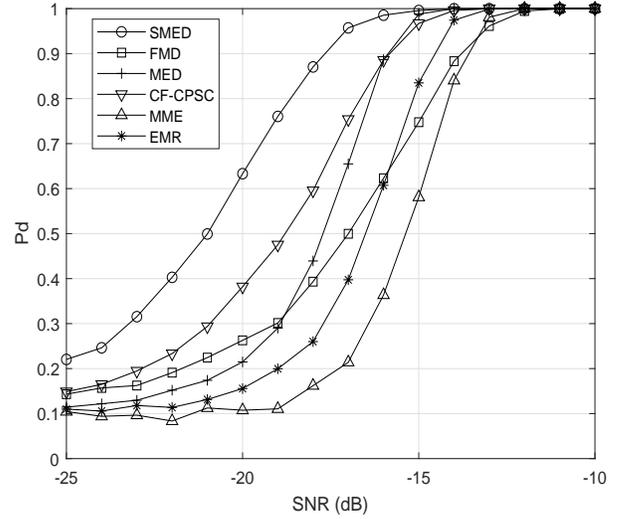}
\caption{Detection probabilities vs. SNR among different algorithms for BPSK signal}\label{fig:pd_snr_n40p100}
\end{figure}
%figure ----------------------------------------------------------

 Figure \ref{fig:pd_snr_n40p100} gives the detection probabilities of different algorithms under SNR. Except for the three eigenvalue detections mentioned above, MME \cite{zeng2009},EMR \cite{huang2015} and CF-CPSC \cite{bomfin2017} algorithms are simulated and compared. These three algorithms are completely blind. Due to the oversampling bandwidth is 10MHz, there is only one subband in CF-CPSC algorithm has signal, so that it obtains a good detection performance since CF-CPSC applies detection by the ratio of subband energy to wholeband energy. With the analysis in above section, the trace method is not suitable for the case of single signal eigenvalue. EMR applies the accumulation of higher-order moment of eigenvalues. Its performance shows just better than that of MME.

\begin{figure}[t]\center
\includegraphics[width=0.5\textwidth, height=0.3\textheight]{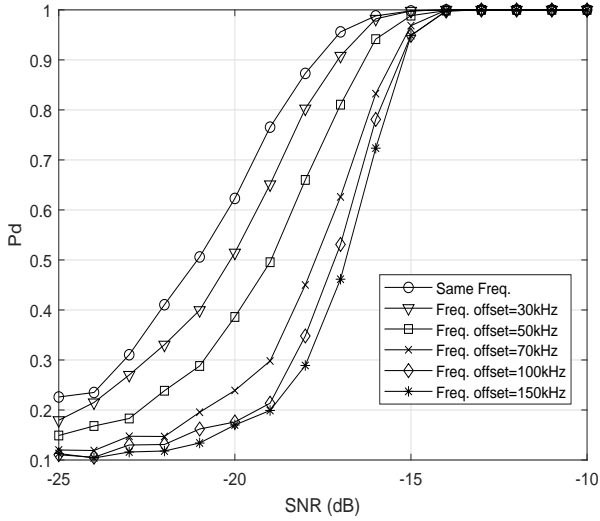}
\caption{SMED performance with different correlative auxiliary signals}\label{fig:vsmed-complex_n40p100_diff}
\end{figure}

 SMED algorithm requires a $priori$ information about the signal to be detected. The information is used to constructing the auxiliary signal with same maximum eigenvector. Thus, the maximum eigenvalue corresponding to the same eigenvector is activated and breaks through M-P law bulk by combining the receiving signal with the auxiliary signal. Therefore, SMED is a semi-blind algorithm. In the simulation, we design the auxiliary signal with the same carrier frequency as the received BPSK carrier. However, we find that this correlationship between the two signals has some tolerance range. In other words, SMED may gain from the auxiliary signal even if its correlationship is weak with the detecting signal. Based on the same parameters as above simulations, other five auxiliary signals, which have frequency offsets from 30kHz$\sim$150kHz, are applied in SMED experiments. Shown in Fig. \ref{fig:vsmed-complex_n40p100_diff}, the $P_d$ of SMED is faded to near MED performance when the frequency offsets of auxiliary signals change from 0 to 70kHz. When the frequency offset further increases, the detection probability changes slightly. Therefore, SMED algorithm has a tolerance range of relationship to the auxiliary signal. Even if this relationship is weak, the detection performance will not be deteriorated significantly.

% figure --------------------------------------------------------
%figure ----------------------------------------------------------

\begin{figure}[t]\center
\includegraphics[width=0.5\textwidth, height=0.3\textheight]{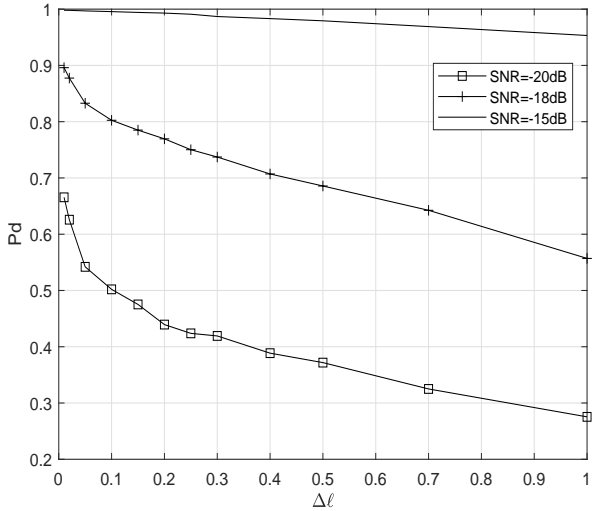}
\caption{optimal amplitudes of auxiliary signal in different SNR environments}\label{fig:pd_vs_diff}
\end{figure}

 Figure \ref{fig:pd_vs_diff} gives the detection probabilities under different amplitudes of auxiliary signal. We measure three different orginal SNRs of received signal, including$-20dB<(1+\sqrt{c}) $ ,$-18dB\approx(1+\sqrt{c}) $ , and$-15dB>(1+\sqrt{c}) $ . Due to unknown SNR in detection, the minimum $\Delta\ell $  is added on the received signal to make $\ell_1^0\rightarrow(1+\sqrt{c})^+ $  under the hypothesis $\mathcal{H}_0 $ . The similar results from simulations with the theoretical results in Fig. \ref{fig:vsopm} are obtained.

% figure --------------------------------------------------------
%figure ----------------------------------------------------------

\begin{figure}[t]\center
 \subfigure[PSD of DTV]{
 \includegraphics[width=0.5\textwidth, height=0.23\textheight]{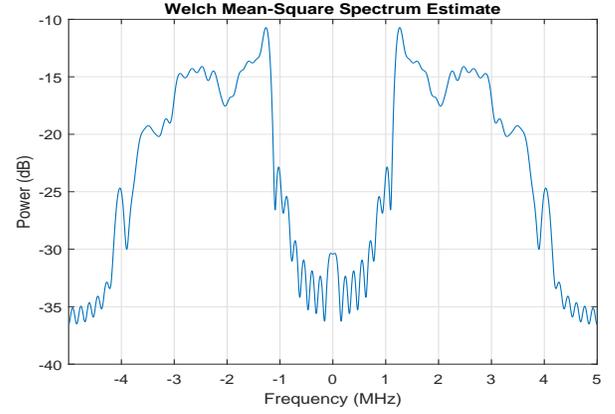}}
 \subfigure[ascending eigenvalues of DTV]{
 \includegraphics[width=0.5\textwidth, height=0.23\textheight]{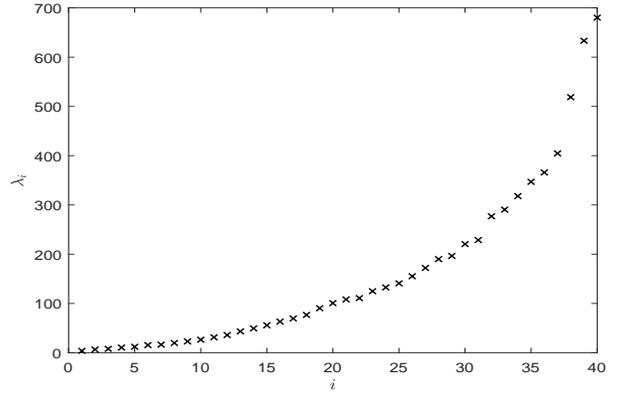}}
 \caption{ properties of the acquired orignal DTV signal}\label{fig:dtv}
 \end{figure}

 For the detection of multi-eigenvector signals, the DTV acquired signal used in cognitive radio is employed \cite{dtv2006}. By combining the complex Gaussian noise with DTV signal, different signal-to-noise ratio signal is simulated. DTV signal has an approximate power spectrum density with 8 MHz spectrum width shown in Fig. \ref{fig:dtv}(a). With the covariance eigenvalue analysis of the original DTV signal, we get the ascending eigenvalues in Fig. \ref{fig:dtv}(b). From the distribution of the eigenvalues, we note that DTV signal is a multiple-eigenvalues and non-single frequency signal.

 % figure --------------------------------------------------------
 % figure --------------------------------------------------------

\begin{figure}[t]\center
\includegraphics[width=0.5\textwidth, height=0.3\textheight]{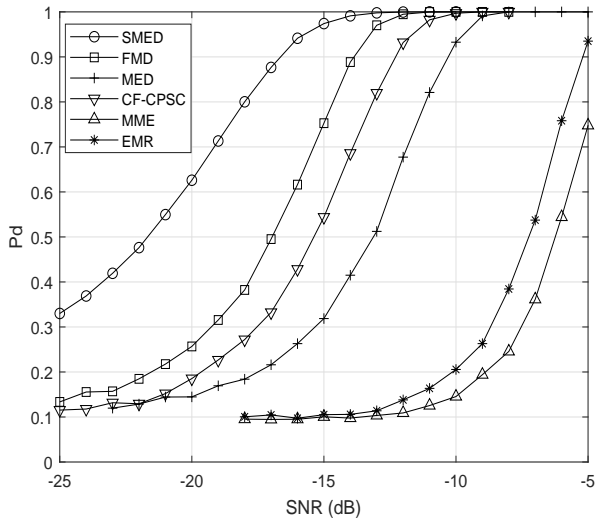}
\caption{Detection probabilities vs. SNR for DTV signal}\label{fig:dtv_pd_n40p100}
\end{figure}

 Suppose that the signal component corresponding to the maximum eigenvalue is prior known. In the simulations, we decompose the orginal DTV signal by SVD, and construct the auxiliary signal with the U and V eigenvector corresponding to the maximum eigenvalue. In practice, this auxiliary signal is able to be obtained by training or searching. Simulation results are shown in Fig.\ref{fig:dtv_pd_n40p100}. SMED algorithm still shows outstanding performance with the help of auxiliary signal. This is mainly attributed to the fact that the magnitude of the maximum eigenvalue is still far greater than that of most eigenvalues even if the signal has multiple eigenvalues. Trace method (FMD) obtained much better performance under DTV signal than under BPSK signal. According to the analysis of Eq. \ref{eq:vartrace}, trace method collects the energy of all the signal components. However, the equal-weight combination of eigenvalues in trace method also results in the sum of variance of all eigenvalues, which reduces the detection performance. From this point of view, the optimal weight combination should be used to obtain the optimal performance like in \cite{liu2017}. Even so, it is difficult to obtain the optimal weights in the low SNR environment in practice. Since the acquired DTV signal is a real signal with symetric double sideband, the frequency domain estimation algorithm of CF-CPSC also gains better performance. It is worth noting that CF-CPSC algorithm requires the fluctuating in PSD shown in Fig. \ref{fig:dtv}(a).

 % figure --------------------------------------------------------
%figure ----------------------------------------------------------

 \section{Conclusion}
In this paper, the eigenvalue detection based SMED signal detection algorithm is proposed, which improves the signal recognition ability by prior correlated auxiliary signal. The performance of SMED algorithm in low SNR environment is verified through theoretical analysis and simulation. SMED algorithm shows good performance not only in single-eigenvector environment but also in wideband and multi-frequency environments. It is not only a spectrum sensing algorithm, but also a universal eigenvalue signal detection algorithm which can obtain high-sensitivity signal detection effects by adding correlative auxiliary signals, and can be widely applied to the fields including signal detection, radar target detection, intrusion detection and fault detection, etc.

% \section*{Acknowledgment}

\ifCLASSOPTIONcaptionsoff
  \newpage
\fi



\bibliographystyle{IEEEtran}

%\begin{biography}[{\includegraphics[width=1in,height=1.25in,clip,keepaspectratio]{mshell}}]{Michael Shell}

% \end{IEEEbiography}



%\vfill


\end{document}